\documentclass[aps,pra,preprint,11pt]{revtex4-2}

\usepackage{amssymb}
\usepackage{amsmath}
\usepackage{bm}
\usepackage{graphics,graphicx}
\usepackage{epsfig,epsf}
\usepackage{latexsym}
\usepackage{epic,eepic,graphicx,indentfirst}
\usepackage{bm}
\usepackage{graphicx}
\usepackage[english]{babel}
\usepackage[latin1]{inputenc}
\usepackage{multirow}
\usepackage{color}
\usepackage[colorlinks={true}]{hyperref}
\hypersetup{citecolor={blue}, filecolor={blue}, linkcolor={blue}, urlcolor={blue}}
\newcommand{\bra}[1]{\langle #1 \vert}
\newcommand{\ket}[1]{\vert #1 \rangle}

\begin{document}

\title{\bf Formation of deeply bound polar molecules combining pump-dump pulses with infrared radiation}

\author{Emanuel Fernandes de Lima}
\email{emanuel@df.ufscar.br}
\affiliation{Departamento de F\'isica, Universidade Federal de S\~ao Carlos (UFSCar)\\ S\~ao Carlos, SP 13565-905, Brazil}
\date{\today}

\begin{abstract}
We consider the formation of cold ground-state polar molecules in a low vibrational level by laser fields. Starting from a pair of cold colliding atoms of dissimilar species, we propose a strategy consisting of three steps. In the first step, a pump pulse induces the molecule formation by photoassociating the atomic pair in a high or intermediate vibrational level of an excited electronic molecular state. This step is followed by a dump pulse in an intermediate vibrational level of the ground state. In the last step, an infrared chirped pulse induces downward transitions among the vibrational levels of the ground electronic state, reaching the ground vibrational level. Initially, the strategy is constructed with fixed-shaped pulses. Subsequently, we perform the calculations with optimized chirped pulses introducing an optimal control technique in which the optimization of the time-dependent frequency is carried out in the time domain. The proposed scheme is an alternative to the use of two pairs of pump-dump pulses or to the direct photoassociation and vibrational stabilization in the ground state. 
\end{abstract}

\maketitle

\section{Introduction}

The controlled formation of molecules from pairs of colliding atoms by the use of external fields is an important endeavor for Physics \cite{PhysRevLett.124.133203,ref1}. It represents not only a very fundamental chemical reaction, but also a major means of creating ultracold molecules, which can be used in many possible applications \cite{doi:10.1146/annurev-physchem-090419-043244,1367-2630-11-5-055049}. In particular, there is a special interest in forming ultracold gases of polar molecules due to their log-range dipole interaction \cite{Chen2023,PhysRevLett.116.205303,PhysRevLett.113.255301,PhysRevLett.113.205301,PhysRevLett.101.133004,NatPhysospelkaus2008,Menegatti2008,PhysRevLett.94.203001,PhysRevLett.92.033004,PhysRevLett.90.043006}. 

Photoassociation reactions, in which the atoms are bind through the interaction of a laser field, constitute one of the main routes leading to the formation of ultracold molecules \cite{PhysRevA.95.013411,PhysRevLett.115.173003,PhysRevLett.105.203001,doi:10.1021/cr300215h,PhysRevA.78.053404,RevModPhys.78.483}. This process is more commonly achieved inducing a transition from the initial pair of atoms to a electronic excited molecular state using visible or ultraviolet radiation. In order to produce stable samples, the photoassociation process needs to be followed by a vibrational stabilization step, aiming at leaving the molecules in the vibrational ground level of the electronic ground state. This second step can be performed by either spontaneous or induced emission \cite{1367-2630-15-12-125028,PhysRevA.82.043437}. A pair of pulses can be employed for this goal, the pump, photoassociating pulse, and the dump, stabilization pulse. However, in order to reach the ground vibrational level efficiently, two sequences of pump-dump pulses may be needed \cite{PhysRevA.98.053411,PhysRevA.103.033301}.

In contrast with photoassociation processes which involve electronic excited states, it is in principle possible to perform both photoassociation and vibrational stabilization solely in the electronic ground state by means of infrared (IR) fields \cite{deLima2015,PhysRevLett.99.073003,Niu20067,deLima200648,Korolkov1996604}. IR fields can induce free-bound and bound-bound transitions due to the existence of a non-negligible permanent dipole moment of colliding atomic pair from distinct species. This approach, though restricted to heteronuclear molecules, has the benefit of not relying on the lifetime and structure of excited states. However, the coupling from the initial continuum levels representing the colliding atomic pair to the bound vibrational levels of the ground electronic are usually small compared to the coupling to bound vibrational levels of excited electronic states. Additionally, even if photoassociation could be achieved within the ground state, the stabilization step through transitions amomg its vibrational levels can be hampered by the existence of forbidden adjacent transitions at certain vibrational levels \cite{PhysRevA.105.013117,de_Almeida_2019}

The formation and vibrational stabilization of the molecules can be enhanced by the temporal shaping of the laser pulses involved in these processes. Optimal quantum control theory has been often invoked to accomplish this task with considerable success, but still with limited experimental implementations. This can be attributed in part to the complexity of the optimized fields. A simpler and more feasible approach is the use of optimized linearly chirped pulses, for which the frequency varies linearly with time \cite{PhysRevA.78.053404,PhysRevA.61.041401,luc-koenig04,PhysRevA.73.033408}. Furthermore, with the increasing technological advances, some progress has been obtained with the use of shaped frequency chirps \cite{doi:10.1021/acs.jpca.5b10088,PhysRevLett.115.173003,Kaufman2017,DELFYETT2012475,0022-3727-44-8-083001}.  In a recent work, we have proposed an analytical form of the frequency of the laser pulse to perform photoassociation along with stabilization in the cold regime \cite{de_Almeida_2019}. However, this approach works only for long pulse duration (tens of nanoseconds) and for low field amplitudes. 

In the present work, we investigate the formation of cold heteronuclear molecules in deeply bound vibrational levels of the ground electronic state from cold colliding atoms. We propose a photoassociation and vibrational stabilization scheme based on three steps. The first two steps are carried out by a pair of pump-dump pulses, which induce the molecular formation in an excited electronic state and a subsequent transition to a intermediate vibrational level of the electronic ground state. These two steps are similar to the stimulated Raman adiabatic passage (STIRAP). The third step employs a chirped IR pulse to perform the final vibrational stabilization to the ground level. This scheme can be an alternative to the use of two pairs of pump-dump pulses and to the IR-photoassociation in the ground state, exploring the strong coupling of the collision atomic pair with excited eletronic state and also performing the stabilization with IR radiation. We apply quantum control theory to optimize the frequency chirping and perform numerical calculations in a model system for the formation of LiCs molecules.

\section{Three-steps scheme for photoassociation}

We envisage a scenario of two cold atoms of different species colliding in the presence of an external laser field. The situation is described by means of two electronic states, the ground state and an electronic excited state, represented by the Hamiltonian

\begin{equation}
\hat{H}(t)=\hat{H}_0-\varepsilon(t)\hat{\mu},
\label{eq:hamiltonian}
\end{equation}
with the molecular Hamiltonian $\hat{H}_0$ given by,

\begin{equation}
\hat{H}_0=
 \begin{pmatrix}
  \hat{T}_g+\hat{V}_g & 0  \\
  0 & \hat{T}_e+\hat{V}_e \\
 \end{pmatrix}
\label{eq:hamiltonianH0}
\end{equation}
where $\hat{T}$ denotes the kinetic energy operator, $\hat{V}$ the potential energy operator, while $\hat{\mu}$ is the dipole moment operator, 

\begin{equation}
\hat{\mu}=
 \begin{pmatrix}
  \hat{\mu}_g & \hat{\mu}_{ge}  \\
  \hat{\mu}_{ge} & \hat{\mu}_e \\
 \end{pmatrix}
\label{eq:dipolematrix}
\end{equation}
where the subscript $g$ refers to the electronic ground state and the subscript $e$ to the excited state. $\hat{\mu}_{ge}$ is the transition dipole moment between the ground and excited electronic states.

The dynamics is given by the time-dependent Schr\"odinger equation,
\begin{equation}
i\hbar\frac{\partial }{\partial t}\ket{\psi(t)}=\hat{H}\ket{\psi(t)},
\label{eq:tdse}
\end{equation}
for the two-component wavefunction,
\begin{equation}
\langle r \ket{\psi(t)}=\begin{pmatrix}
	\psi_g(r,t) \\
	\psi_e(r,t)
\end{pmatrix},
\label{eq:tdwf}
\end{equation}
where $r$ is the interatomic distance. In order to capture the essential physics of the problem, we retain only the vibrational motion, neglecting the rotational motion, due to the low temperatures involved. 

\begin{figure}[ht!]%
\includegraphics[width=1\textwidth]{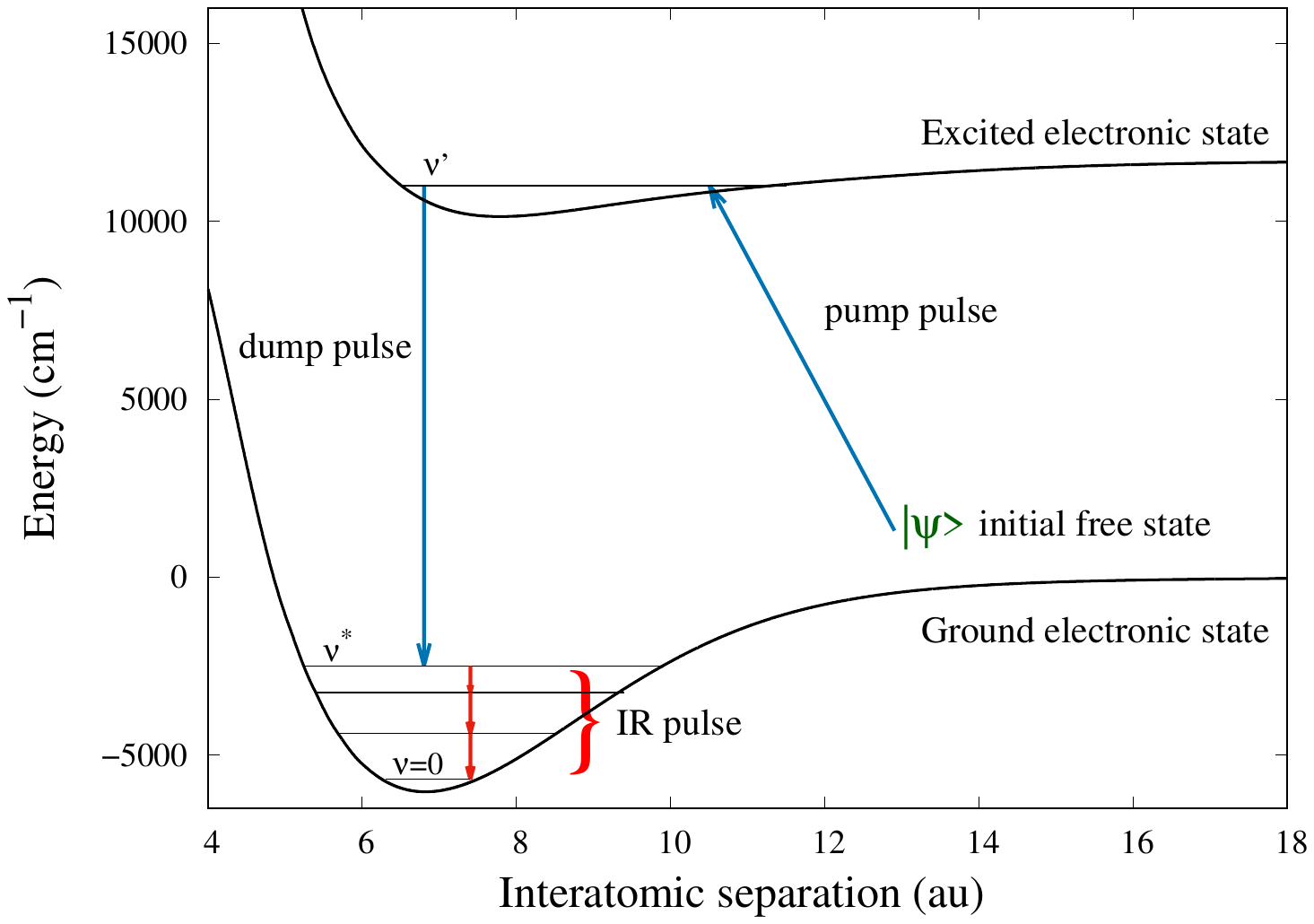}
\caption{(color online) Schematic diagram of the three-step photoassociation along with vibrational stabilization. In the first step, the pump pulse drives the initial unbound state to a vibrational level $\nu'$ of an excited electronic state. In the second step, the pump pulse drives a transition to an excited vibrational level $\nu^*$ of the electronic ground state. In the last step, an IR pulse performs further vibrational stabilization within the ground electronic state by driving downward transitions and reaching the vibration ground level $\nu=0$.}\label{3Scheme}
\end{figure}

The external time-dependent electric field $\varepsilon(t)$ is composed of three pulses,
\begin{equation}
\varepsilon(t)=\varepsilon_p(t)+\varepsilon_d(t)+\varepsilon_{\rm ir}(t),
\label{eq:externalfield}
\end{equation}
where $\varepsilon_p(t)$ refers to the pump pulse, $\varepsilon_d(t)$ refers to the dump pulse and $\varepsilon_{\rm ir}(t)$ refers to infrared pulse. Figure~\ref{3Scheme} sketches the three-step scheme for forming polar molecules in their ground vibrational level. First, the initial unbound state $\ket{\psi}$, which represents the colliding atomic pair, is driven to a vibrational level $\nu'$ of an excited electronic state by a pump pulse. In the next step, the transition between $\nu'$ and an excited level $\nu^*$ of the ground electronic state is induced by a dump pulse. Finally, the last step consist of a vibrational stabilization step, where the transition from $\nu^*$ to the ground level $\nu=0$ is induced by a single chirped IR pulse.

\pagebreak

\section{Model for the LiCs molecule}

The energy spectrum of each molecular potential is composed of a set of discrete levels and a continuum region. As practical numerical solution of the dynamical equations often requires, we consider the continuum part is discretized, for instance, by the addition of an infinite barrier at long range, or equivalently, by the introduction of a spatial grid \cite{1751-8121-41-33-335303}. We label the discrete energy levels of the ground state by $E_{\nu}$ with corresponding eigenfunctions $\ket{\phi_{\nu}}$ and the discretized continuum levels by $E_n$ with corresponding eigenfunctions $\ket{\phi_{n}}$. The associated energy and eigenfunctions of the excited state are distinguished from those of the ground state by a prime symbol.

We illustrate the three-steps scheme considering the ${\rm X}^1\Sigma$ and ${\rm B}^1\Pi$ states of LiCs as the ground and excited potentials $V_g(r)$ and $V_e(r)$, respectively. The potential energy functions and dipoles couplings were obtained from spline fittings based on data available in the literature \cite{:/content/aip/journal/jcp/140/18/10.1063/1.4875038,doi:10.1021/jp101588v,:/content/aip/journal/jcp/131/5/10.1063/1.3180820,staanum2007,aymar2005}. In order to obtain the vibrational energies $E_{\nu}$ and eigenfunctions $\ket{\phi_{\nu}}$ of each electronic state, we have resorted to the B-Splines basis set with an exponential sequence of breakpoints \cite{masnou2009,0034-4885-64-12-205}. We have obtained $51$ vibrational bound levels for the ground state and $34$ bound levels for the excited state. In this approach, the continuum eigenfunctions are discretized in a large box of size $r_{\rm max}$ (we have used $r_{\rm max}$ typically of the order of $1500$ atomic units).

\begin{figure}[ht!]%
\includegraphics{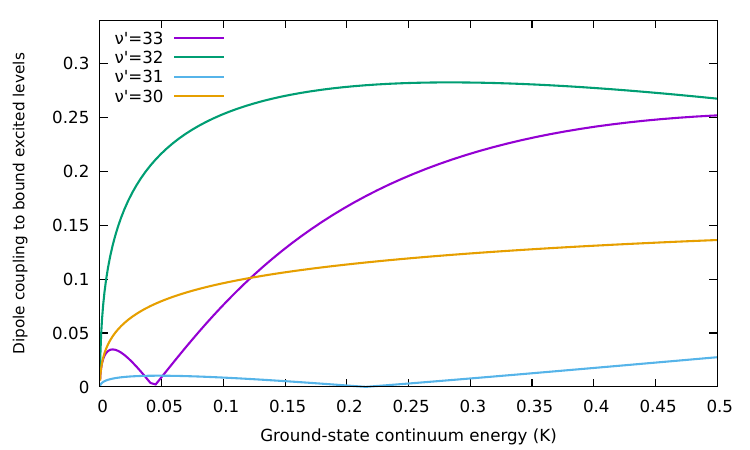}
\caption{(color online) Dipole coupling between the free levels of the ground state and the bound levels of the excited state.}\label{DisContge}
\end{figure}

\begin{figure}[ht!]%
\includegraphics{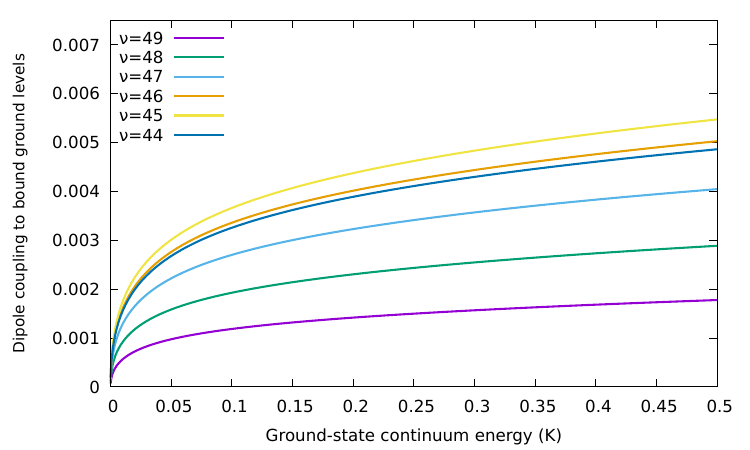}
\caption{(color online) Dipole coupling between the free levels of the ground state and its bound levels.}\label{DisContgb}
\end{figure}

Figure~\ref{DisContge} shows the absolute value of the transition dipole moment  $|\bra{\phi_{\nu'}}\mu_{ ge}\ket{\phi_n}|$ from the continuum levels of the ground-electronic state to some top bound levels of the excited electronic state. The level $\nu'=32$ possess the highest coupling up to the temperature $T=0.5$K, revealing a possible starting point for the photoassociation process. We note that the lower vibrational levels of the excited state (not shown in the figure) have even smaller coupling. Figure~\ref{DisContgb} shows the dipole coupling between the continuum levels and the bound levels of the ground state. An available route is for IR ground-state photoassociation is a transition to the $\nu=45$ level, which has the highest coupling. However, these coupling are generally two orders of magnitude lower than the corresponding couplings to the excited bound levels.

\begin{figure}[ht!]%
\includegraphics{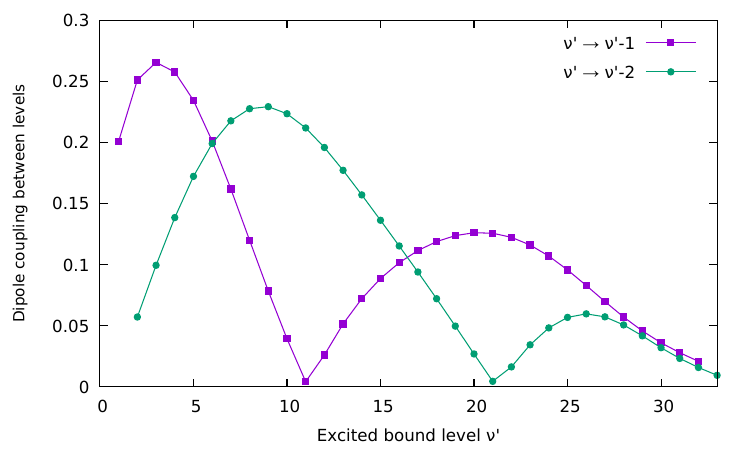}
\caption{(color online) Dipole coupling between the bound levels of the excited state for one and two photons transitions.}\label{DisDisee}
\end{figure}

\begin{figure}[ht!]%
\includegraphics{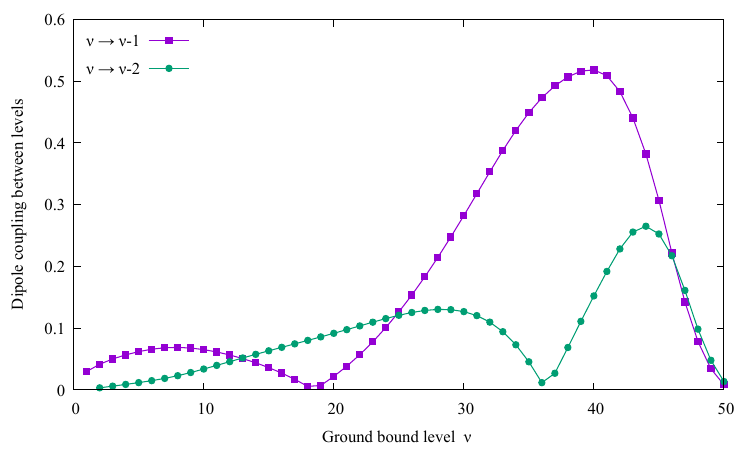}
\caption{(color online) Dipole coupling between the bound levels of the ground state for one and two photons transitions.}\label{DisDisgg}
\end{figure}

Figure~\ref{DisDisee} shows the dipole couplings between bound levels of the excited state corresponding to one and two-photons downward transitions. We note that direct vibrational stabilization in the excited state is possible, but there are some very small couplings for the one photon (at $\nu'=11$) and two photons transitions (at $\nu'=21$), which should difficult the descending from the vibrational levels. However, one can avoid these so-called "missing rugs", by switching from one-photon to two-photons transition frequencies. A similar scenario is observed for the downward transitions within the ground state, as shown in Fig.~\ref{DisDisgg}. The gaps for the transitions are noted around $\nu=19$ (one-photon) and $\nu=36$ (two-photons). In contrast to the excited state, the coupling are larger between the top vibrational levels.

\begin{figure}[ht!]%
\includegraphics{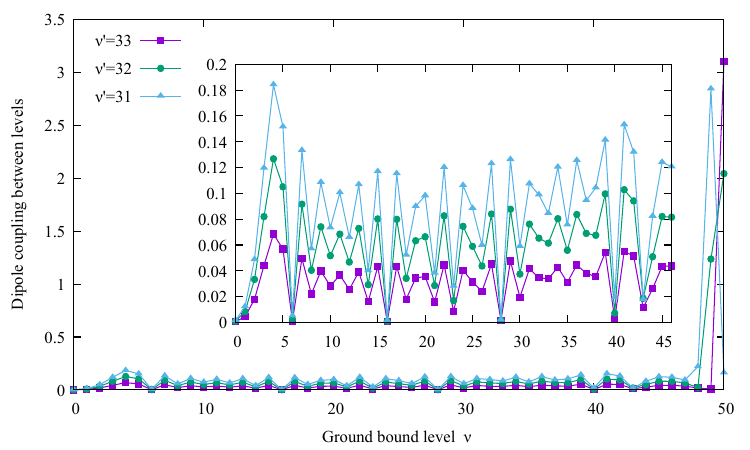}
\caption{(color online) Transition dipole moment from top levels of the electronic excited state to the bound levels of the ground state.}\label{DisDiseghigh}
\end{figure}

Figure~\ref{DisDiseghigh} shows the absolute values of the transition dipole moments among the top levels of the excited state and the bound levels of the ground state. It can be noted that these top levels have a very strong coupling with the very top levels of the ground state. Nevertheless, as shown in the inset, they still have considerable couplings to the $\nu=4$ level. Thus, once the excited level $\nu'=32$ is populated, it can be transferred to the $\nu=4$ state by the action of the dump pulse.

\pagebreak

\pagebreak

\pagebreak

\section{Fixed-shape pulses}

As seen in Figs.~\ref{DisContge} and \ref{DisContgb}, for low temperatures, the coupling of the continuum with the bound vibrational levels indicates that the more favored route is to form a molecule in the $\nu'=32$ level of the excited electronic state. This transition will be induced by the pump pulse. As noted in Fig.~\ref{DisDiseghigh}, there is a considerable coupling between the $\nu'=32$ with the $\nu=4$ of the ground state, which is induced by the dump pulse. Subsequently, the IR chirped pulse drives the transition to the ground level.

We consider the three-steps procedure with fixed shaped pulses, 
\begin{equation}
\varepsilon_i(t)=S_i(t)A_i\sin\left[\alpha_i(t)\right], \;\;i=p,d,{\rm ir}
\label{eq:externalfield}
\end{equation}
where $A_i$ is the peak amplitude of the ith-pulse, while the envelope $S_i$ is given by,
\begin{equation}
S_i(t)= \begin{cases}
\sin^2\left(\frac{\pi}{2}\frac{t-t_0^i}{t_u^i-t_0^i}\right) & \text{, if } t_0^i<t<t_u^i \\
1 & \text{, if } t_u^i<t<t_d^i \\
\sin^2\left(\frac{\pi}{2}\frac{t+t_w^i-2t_d^i}{t_w^i-t_d^i}\right) & \text{, if } t_d^i<t<t_w^i \\
0 & \text{, otherwise,}
\end{cases}
\label{eq:envelope}
\end{equation}
where $t_w^i-t_0^i$ defines the duration of the pulse that starts at $t=t_0^i$, while $t_u^i-t_0^i$ and $t_w^i-t_d^i$ set the times of smooth switching on and off, respectively.

The pump and dump pulses are chosen to be unchirped pulses, $\alpha_p=\omega_pt$ and $\alpha_d=\omega_dt$, while the IR pulse has a linear chirping rate,
\begin{equation}
    \alpha_{\rm ir}=\omega_{\rm ir}t+\frac{1}{2}\chi t^2
\end{equation}
 
 We have performed numerical calculations taking the initial wavefunction representing the atomic collision as a Gaussian wavepacket in the electronic ground state,
\begin{equation}
\langle r \ket{\psi(t=0)}=\begin{pmatrix}
	\left(\frac{2}{\pi a^2}\right)^{1/4}\exp\left[{\rm i}\kappa r-\frac{\left(r-r_0\right)^2}{a^2}\right] \\
	0
\end{pmatrix},
\label{eq:iniwf}
\end{equation}
where $a$ and $r_0$ define respectively the initial width and the central position of the wavepacket, while $\kappa<0$ sets the initial collision momentum. For the time propagation, we have considered an initial state with position $r_0=450$a.u., width $a=100$a.u., and with collision energy corresponding to $50$mK. This wavepacket is placed far from the interaction region, i.e., at $t=0$ the potentials and dipole functions are negligible over the wavepacket range.

In order to solve the time-dependent Schr\"odinger equation, we have expanded the wavefunction in the basis of the energy eigenfunctions, truncating the number of discretized continuum levels for each electronic state. Substituting the expansion in Eq.~(\ref{eq:tdse}) yields a coupled system of first-order differential equations for the expansion coefficients. The corresponding unitary propagator $U(t,0)$ is then written in terms of a series of short time step propagators $U(t+\Delta t,t)$ and each short-time propagator is in turn approximated by a second-order split-operator, leading to the evolution of the wavefunction $\ket{\psi(t+\Delta t)}=U(t+\Delta t,t)\ket{\psi(t)}$ (for details, see Refs. \cite{deLima2015,deLima2011267}). The number of discretized continuum levels was typically $500$ for the ground state and $200$ for the excited state.

\begin{figure}[ht!]%
\includegraphics{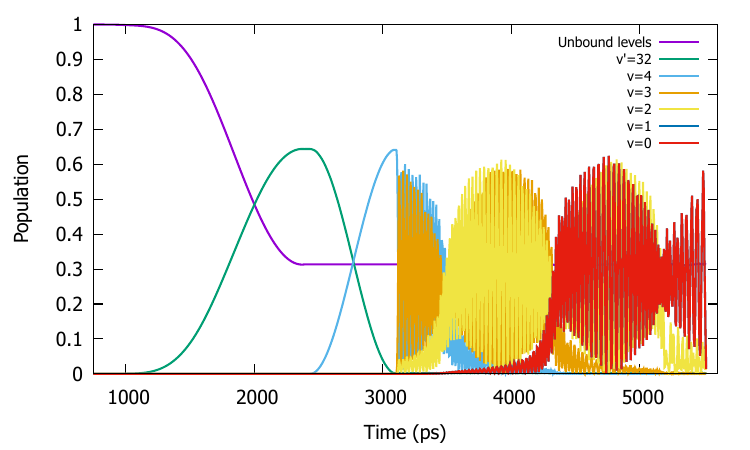}
\caption{(color online) Population dynamics of the three-step scheme with fixed-shape pulses. The unbound population corresponds to the initial state.}\label{popd_3steps}
\end{figure}

Figure~\ref{popd_3steps} illustrates the three steps scheme with fixed-shape pulses. It is possible to notice that obtain high yields in the ground level $\nu=0$ (about $60\%$). However, the intensity of the last IR chirped pulse is quite large (of the order of hundreds of $MVcn^{-1}$), which causes the fast oscillation of the populations above $t=3$ns. As we show in the next section, the optimization of the chirping can perform the stabilization process with considerable lower intensity. 

\section{Optimized chirped pulses}

Consider the control problem of transferring some initial state $\ket{\psi_0}$ to a target vibrational level $\ket{\phi_{\nu}}$ in a time $t=t_f$ with a single pulse of the form of Eq. (\ref{eq:externalfield}), where $\alpha(t)$ is the control function to be determined (for simplicity, we will drop here the index $i$). To this end, we define the extended functional $J$ to be maximized,

\begin{equation}
J[\alpha]=\left|\bra{\phi_{\nu}}\left.\psi(t_f)\right>\right|^2-2\Re\left\{\int_0^{t_f}\bra{\chi(t)}\frac{\partial}{\partial t}+i\hat{H}\left[\alpha(t)\right]\ket{\psi(t)}\right\},
\label{eq:func1}
\end{equation}
where $\ket{\psi(t)}$ is the wavefunction driven by the pulse and $\ket{\chi(t)}$ is the Langrange multiplier introduced to assure satisfaction of the Schr\"odinger equation. The necessary condition for optimality, $\delta J=0$, yields the equations to be satisfied by the wavefunction, Lagrange multiplier and optimal frequency,

\begin{equation}
i\frac{\partial}{\partial t}\ket{\psi(t)}=\hat{H}\left[\alpha(t)\right]\ket{\psi(t)},
\label{eq:SEq}
\end{equation}
with $\ket{\psi(0)}=\ket{\psi_0}$,

\begin{equation}
i\frac{\partial}{\partial t}\ket{\chi(t)}=\hat{H}\left[\alpha(t)\right]\ket{\chi(t)},
\label{eq:LEq}
\end{equation}
with $\ket{\chi(t_f)}=\ket{\phi_{\nu}}\left<\phi_{\nu}\right.\ket{\psi(t_f)}$ and

\begin{equation}
\Im\left\{\bra{\chi(t)}\hat{\mu}\ket{\psi(t)}\right\}S(t)A\cos\left(\alpha(t) \right)=0.
\label{eq:FEq}
\end{equation}
Any $\alpha(t)$ which satisfies the above three coupled equations is a local optimal solution to the control problem. In order to numerically solve the optimal control problem, we apply the TBQCP scheme, in which $\alpha(t)$ is updated in the $n$-th iteration according to,

\begin{equation}
\alpha^{n+1}(t)=\alpha^{n}(t)-\eta_0f_{\mu}^{n}(t),
\label{eq:tbqcp}
\end{equation}
where $\eta_0$ is a constant to be numerically adjusted for convergence and $f_{\mu}$ is the gradient $\delta J/\delta \alpha$,
\begin{equation}
f_{\mu}^{n}(t)=-2\Im\left\{\bra{\chi^n(t)}\hat{\mu}\ket{\psi^n(t)}\right\}S(t)A\cos\left(\alpha^{n}(t) \right).
\label{eq:grad}
\end{equation}
The algorithm is started by choosing a trial control $\alpha^{0}(t)$ and solving the equation for the Lagrange multiplier (\ref{eq:LEq}). Then, the gradient is computed as the wavefunction is propagated forward in time.

\begin{figure}[ht!]%
\includegraphics{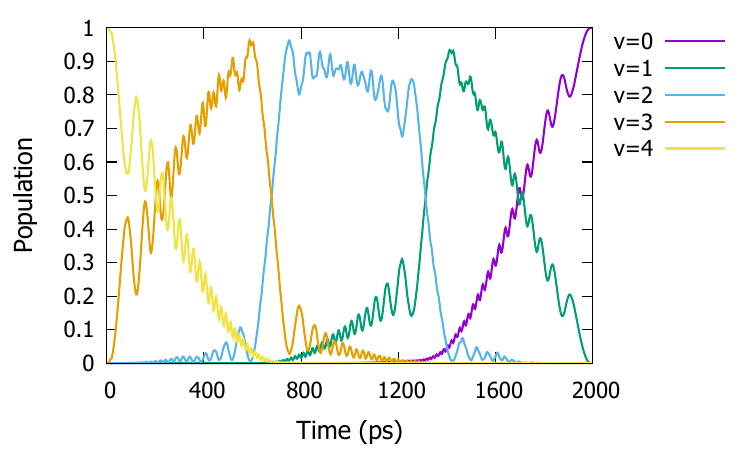}
\caption{(color online) Population dynamics of the IR vibrational stabilization process with the optimized chirped pulse.}\label{popd_optchirp}
\end{figure}

\begin{figure}[ht!]%
\includegraphics{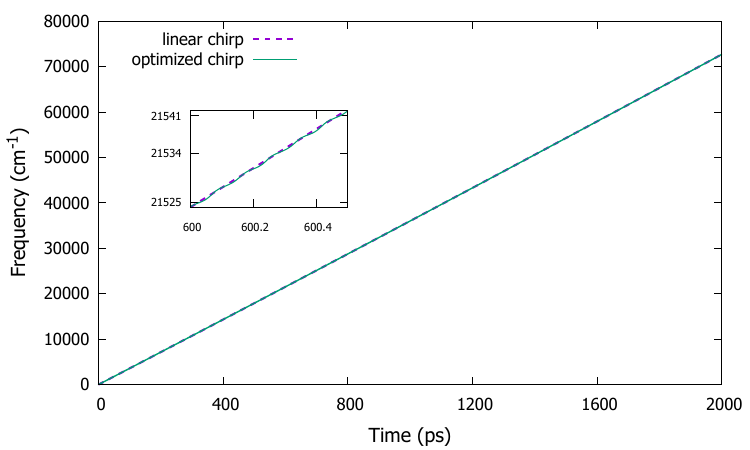}
\caption{(color online) Comparison of the linear chirp with the optimized chirp. The inset shows the same in a small time interval.}\label{optchirp}
\end{figure}

Figure~\ref{popd_optchirp} shows the resulting population dynamics with the optimized chirped pulse build to perform the vibrational stabilization from the $\nu=4$ to the ground level $\nu=0$. We note that complete transfers can be achieved with the optimized pulse despite of its comparative low amplitude (of the order of few $\rm MVcm^{-1}$). Figure~\ref{optchirp} compares the optimized chirped frequency with the linear chirp. At first sight, there is no difference between the two chirped frequencies. But, as the inset shows, the optimum chirp has a modulation about the linear chirp.

\pagebreak

\section{Conclusions}

In this work, we propose the combination of a pair of pump-dump pulses with a chirped IR pulse to perform the photoasociation of cold atoms along with the vibrational stabilization of the associated cold polar molecule. The proposed three-pulses scheme takes advantage of the high coupling of the initial collision state to an excited electronic states as compared to the electronic ground state. By performing part of the vibrational stabilization in the selected excited electronic state, it can avoid "missing rung" problems implied in the ground-state-only stabilization. On the other hand, the scheme populates only a single electronic excited state, avoiding complications of two pairs of pump-dump schemes, which usually involves two excited states. We have also performed the optimization of the chirped IR pulse in time domain, obtaining high yields in the ground vibrational level. Our calculations show that the optimized chirped pulses differs from a linear chirped pulse only by a small modulation of the time-dependent frequency. Therefore, our proposed methodology can be an alternative pathway in order to perform both photoassoiciation and vibrational stabilization aiming at forming cold/ultracold polar molecules.

\section*{Acknowledgments}\label{sec7}
EFL acknowledges support from S\~ao Paulo Research Foundation, FAPESP (grants 2023/04930-4 and 2014/23648-9).


\end{document}